\documentclass[pra,floatfix,preprint,nofootinbib,eqsecnum,12pt]{revtex4}

\usepackage{graphicx}
\usepackage{epstopdf}
\usepackage{amsmath}
\usepackage{enumitem}

\numberwithin{equation}{section}
\renewcommand{\theequation}{\arabic{section}.\arabic{equation}}

\makeatletter
\newcounter{manualsubequation}
\renewcommand{\themanualsubequation}{\alph{manualsubequation}}
\newcommand{\startsubequation}{%
  \setcounter{manualsubequation}{0}%
  \refstepcounter{equation}\ltx@label{manualsubeq\theequation}%
  \xdef\labelfor@subeq{manualsubeq\theequation}%
}
\newcommand{\tagsubequation}{%
  \stepcounter{manualsubequation}%
  \tag{\ref{\labelfor@subeq}\themanualsubequation}%
}
\let\subequationlabel\ltx@label
\makeatother

\begin{document}

\newcommand{\etal}{{\it et al.}\/}
\newcommand{\gtwid}{\mathrel{\raise.3ex\hbox{$>$\kern-.75em\lower1ex\hbox{$\sim$}}}}
\newcommand{\ltwid}{\mathrel{\raise.3ex\hbox{$<$\kern-.75em\lower1ex\hbox{$\sim$}}}}
\renewcommand{\thefootnote}{\fnsymbol{footnote}}

\title{Classical Physics and Hamiltonian Quantum Mechanics \\ as
Relics of the Big Bang\protect\footnote{This paper is based in an Invited talk at Nobel Symposium no. 79: The Birth and Early Evolution of our Universe, Graftavallen, Sweden, June 11-16, 1990. It was published in the proceedings of that conference as {\sl Phys. Scripta} , T36, 228, 1991. The original pdf has been reset in latex with only some typos fixed and some infelicities of expression corrected, and  then posted to arXiv to make it more accessible. Some effort has been made to update the terminology for example replacing `initial condition' with `quantum state'. But no  effort has been made to improve the exposition or to connect to later work.  The references have not been updated.}}

\author{James B. Hartle}
\affiliation{Department of Physics, University of California, Santa Barbara, CA 93106-9530, USA}

\date{\today}

\begin{abstract}
In a fundamental formulation of the quantum mechanics of a closed system
such as the universe as a whole, three forms of information are needed to
make predictions for the probabilities of  alternative time histories of the closed system . These are the
action functional of the elementary particles, the quantum istate 
of the universe, and the  description of  our specific history. We discuss
the origin of the ``quasiclassical realm" of familiar experience and Hamiltonian
quantum mechanics with its preferred time in such a formulation of
quantum cosmology. It is argued that these features of the universe are not
general properties of quantum theory, but rather approximate features that are 
emergent after the Planck time as a consequence of theories of the closed system's quantum state and dynamics.

\end{abstract}


\maketitle


\section{Introduction}\label{sec:1}

An unavoidable inference from the physics of the last sixty
years is that we live in a quantum mechanical universe - a
universe in which the process of prediction conforms to that
framework we call quantum mechanics on all scales from
those of the elementary particles to those of the universe
itself. We perhaps have little direct evidence of peculiarly
quantum phenomena on very large and even familiar scales
today, but there is no evidence that the phenomena that we
do see cannot be described in quantum mechanical terms and
explained by quantum mechanical laws. In the earliest
universe we have a physical
system of the largest possible size for which a quantum
mechanical description is  essential, especially at
times earlier than the Planck time. The nature of this quantum
mechanical description and its observable consequences are
the subject of quantum cosmology.

The most general objects of prediction in quantum
mechanics are the probabilities of alternative histories of the
universe. Three forms of information are needed to make
such predictions. These are the action functional of the
elementary particles, the quantum state of the universe,
and the information that we  already know of about  our past history on which
probabilities for future prediction are conditioned. These are
sufficient for every prediction in science and there are no
predictions which do not  involve
all three forms of information at this fundamental level. A theory of the action functional
of the fundamental fields is the goal of elementary
particle physics. The equally fundamental, equally necessary,
theory of the  quantum state  of the universe \cite{ref1}which is the  goal  of quantum cosmology.  In a final theory these may even be related objectives.

To make contact with observation, a theory of the quantum 
state must explain correlations among observations
today. What are these observations? On
the largest scales these include  the familiar features of the universe whose
explanation cosmology has usually traced to the quantum state. These include the approximate homogeneity and
isotropy, the approximate spatial flatness, the simple spatial
topology, and the spectra of deviations from exact homogeneity
and isotropy which we can see today as large scale
structure and earlier as anisotropies in the background
radiation. On very small scales, 
coupling constants of the elementary particles could be quantum
probabilistic with a probability distribution that may
depend, in part, on the quantum state.  In this talk, however,
I shall discuss two much more familiar features of the universe,
that are accessible at ordinary scales  and   owe their
origin, at least in part, to the quantum state. These
are the applicability of classical physics over much of the late
universe, including especially the existence of classical
spacetime, and the applicability of the Hamiltonian 
quantum mechanics. I shall argue that, in a fundamental
formulation of the quantum process of prediction, these
familiar features of the universe call for explanation just as
much as do the large scale structural characteristics alluded to
above, although there may be a very wide range of quantum states  of the universe from which they follow.

For a quantum mechanical system to exhibit classical
behavior there must be some restriction on its state and some
coarseness in how this classical behavior is described. This is clearly illustrated in
the quantum mechanics of a single particle moving in one dimension. $x$. Ehrenfest's
theorem shows that generally
\begin{equation}
  M\frac{d^2\langle{x}\rangle}{dt^2}=
	\langle-\frac{\partial V}{\partial x}\rangle.\label{eq:1.1}
\end{equation}
However, only for special states, typically narrow wave
pickets, will this become an equation of motion for $\langle x\rangle$ of
the form
\begin{equation}
  M\frac{d^2\langle{x}\rangle}{dt^2}=
	\frac{\partial V(\langle x\rangle)}{\partial x}.\label{eq:1.2}
\end{equation}
For such special states, successive observations of position in
time will exhibit the classical correlations predicted by the
equation of motion (\ref{eq:1.2}) provided that these observations are
coarse enough so that the properties of the state which allow
(\ref{eq:1.2}) to replace the general relation (\ref{eq:1.1}) are not affected by
these observations. An exact determination of position, for
example, would yield a completely delocalized wave packet
an instant later and (\ref{eq:1.2}) would no longer be a good approximation
to (\ref{eq:1.1}). Thus, even for large systems, and in particular
for the universe as a whole, we can expect classical
behavior only for certain  quantum states and then only when a
sufficiently coarse grained description is used. If classical
behavior is in general a consequence only of a certain class of
states in quantum mechanics, then, as a particular case, we
can expect to have classical spacetime only for certain states
in quantum gravity. The classical spacetime geometry we see
all about us in the late universe is not property of every state
in a theory where geometry fluctuates quantum mechanicalIy.
Rather it is  traceable fundamentally to restrictions on the quantum 
state. Such restrictions are likely to be generous in that,
as in the single particle case, many different states will exhibit
classical features. The existence of classical spacetime and the
applicability of classical physics are thus not likely to be very
restrictive conditions on constructing a theory of the quantum 
state. However, they are such manifest and accurate
features of the late universe that it is important to understand
quantitatively the class of quantum states  with which they
are consistent. Any quantum state incorporated into the  theory must
lie in this class.

A feature of the late universe which is closely related to
the existence of classical spacetime is the applicability of
Hamiltonian quantum mechanics. Time plays a special role in
the familiar Hamiltonian formulation of quantum mechanics,
Time is the only observable for which there are no
interfering alternatives like position is an interfering alternative
for momentum. Time is the only observable not represented
in the formalism as an operator but rather enters the theory
as a parameter describing evolution. Thus, just for its formulation,
Hamiltonian quantum mechanics requires a fixed
background spacetime to supply the preferred time. This we
can expect only for special states of the universe and then
only approximately at certain times and in certain regions.

Classical physics is applicable over a wide domain in the
universe. Here in the late universe the geometry of spacetime,
viewed sufficiently coarsely, is classical, definite and evolving
by Einstein's equation. Hamiltonian quantum mechanics
with its preferred time is correct for field theory on all
accessible scales. The point of view I shall describe in this talk
is that these familiar, homey, features of our world are most
fundamentally seen not as exact properties of the basic theory
but rather as approximate, emergent properties of the late
universe appropriate to the particular quantum state  which
our universe does have. Put more crudely, I shall argue that
the classical realm with its classical spacetime and the
Hamiltonian form of quantum mechanics with its preferred
time are {\it relics of the big bang}.

To exhibit the classical realm and Hamiltonian quantum
mechanics as emergent features of the universe we need a
generalization of the Copenhagen framework for quantum
mechanics on at least two counts. First, the various Copenhagen
formulations of quantum mechanics characteristically
posited the existence of the classical realm (or a classically
behaving ``observer") as an additional assumption beyond
the framework of wave function and Schr\"odinger equation
that was necessary to interpret the theory. Second, these
Copenhagen formulations assumed the preferred time of
Hamiltonian quantum mechanics. What has been assumed by theory we
cannot expect to be explained by that theory. A more general framework is
necessary.

In this talk I shall describe some routes towards these
necessary generalizations. The generalizations I shall describe
stress the consistency of probability sum rules as the primary
criterion for assigning probabilities to histories rather than
any notion of ``measurement". They stress the quantum state 
of the universe as the ultimate origin within quantum
mechanics of the classical realm. They stress the sum-over-histories
formulation of quantum mechanics as a potentially
more general and generally covariant framework for a
quantum mechanics of spacetime. The work on these generalization
cannot be said to be complete but the directions
seem promising to me. To keep the discussion manageable I
shall take the discussion in two steps. First, in Sections \ref{sec:2}--\ref{sec:4},
I shall neglect gross fluctuations in the geometry of spacetime;
later, in Sections \ref{sec:5} and \ref{sec:6}, I shall return to the generalization
needed to accommodate them.

\section{Post-Everett quantum mechanics}\label{sec:2}

We begin with a brief review of the post-Everett formulation
of the quantum mechanics of closed systems such as the
universe as a whole.\footnote{Today this would be called consistent or decoherent histories quantum theory}
 As described above, we shall first assume
a fixed background spacetime which supplies a preferred
family of timelike directions. This, of course, is an excellent
approximation on accessible scales for times later than
a very short time after the big bang. The familiar apparatus of Hilbert
space, states, Hamiltonian and other operators may then be
applied to process of prediction. Indeed, in this context, the
quantum mechanics of cosmology is in no way distinguished
from the quantum mechanics of a large isolated box, perhaps
expanding, but containing both the observed and its
observers.

The quantum mechanical framework that I shall describe
has its origins in the work of Everett \cite{ref:2} and has been
developed by many. In its recent developments it incorporates
ideas of Zeh \cite{ref:3}, Joos and Zeh \cite{ref:4}, Zurek \cite{ref:5},
Griffiths \cite{ref:6}, and Omn\`es \cite{ref:7}. The particular development
I shall follow is due to Murray Gell-Mann and myself \cite{ref:8, ref:9}.

\begin{figure}[htbp]
  \includegraphics[width=12cm]{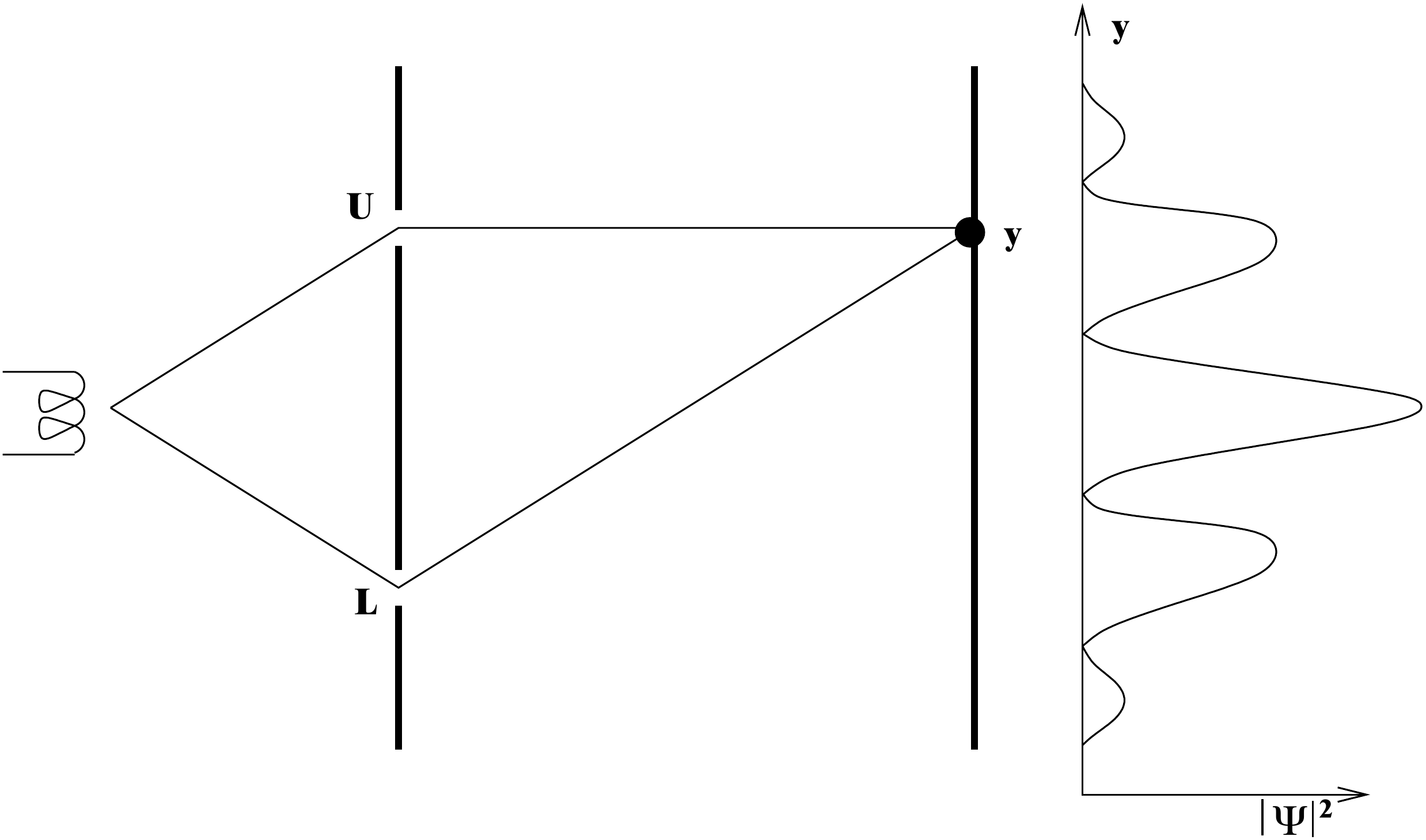}
  \caption{The two-slit experiment. An electron gun at left emits an electron
traveling towards a screen with two slits, its progress in space recapitulating
its evolution in time. When precise detections are made of an ensemble of
such electrons at the screen it is not possible, because of interference, to
assign a probability to the alternatives of whether an individual electron went
through the upper slit or the lower slit. However, if the electron interacts with
apparatus that measures which slit it passed through, then these alternatives
decohere and probabilties can be assigned.
	\label{fig:1}}
\end{figure}

A characteristic feature of a quantum mechanical theory is
that not every history which can be described can be assigned
a probability. Nowhere is this more clearly illustrated than in
the two-slit experiment (Fig.~\ref{fig:1}). In the usual discussion, if we
have not measured which slit the electron passed through on
its way to being detected at the screen, then we are not
permitted to assign probabilities to these alternative histories.
It would be inconsistent to do so since the correct probability
sum rules would not be satisfied. Because of interference, the
probability to arrive at $y$ on the detecting screen  is not the sum of the probabilities
to arrive at $y$ going through the upper and the lower slit because in quantum theory probabilities are squares of amplitudes and 
\begin{equation}
  p(y)\ne p_U(y)+p_L(y)\label{eq:2.1}
\end{equation}
because
\begin{equation}
  |\psi_L(y)+\psi_U(y)|^2\ne |\psi_L(y)^2|+|\psi_U(y)^2|.\label{eq:2.2}
\end{equation}
If we {\it have}\/ measured which slit the electron went through,
then the interference is destroyed, the sum rule obeyed, and
we can meaningfully assign probabilities to these alternative
histories.

It is a general feature of quantum mechanics that a rule is
needed to determine which histories can be assigned probabilities.
As the two-slit example illustrates, in the  standard Copenhagen
formulation that is found in most textbooks, probabilities are assigned to histories which are
{\it measured}\/. This is a rule which assumes a division of the
universe into one subsystem which is measured or observed
and another which does the measuring or observing. Further,
to define measurement, the Copenhagen formulations had, in
one way or another, to posit as fundamental the classical
world that we see all about us. We can have none of this in
cosmology. In a theory of the whole thing there can be no
fundamental division into observer and observed. There is no
fundamental reason for a closed system to exhibit classical
behavior generally in any variables. Measurements and
observers cannot be fundamental notions in a theory which
seeks to describe the early universe where neither existed. We
need a more general quantum mechanics for assigning probabilities to histories in
quantum cosmology.

I shall now describe the rules which specify which histories
of a closed system may be assigned consistent probabilities in
the post-Everett formulation and what these probabilities
are. They are essentially the rules of Griffiths \cite{ref:6} further
developed by Omn\`es \cite{ref:7} and independently but later arrived
at by Gell-Mann and the author \cite{ref:8}. The idea is simple:
Probabilities can be assigned to those coarsely described
histories for which the probability sum rules are obeyed as a
consequence of the particular quantum state the closed system
does have.

To describe the rules in detail, it is convenient to begin
with Feynman's sum-over-histories formulation of quantum
mechanics since histories are our concern. There, all quantum
amplitudes are expressed as functionals of completely fine-grained
histories specified by giving a set of generalized
coordinates $q^i(t)$ as functions of time. These might be the
values of fundamental fields at different points of space, for
example.

Completely fine-grained histories cannot be assigned
probabilities; only suitable coarse-grained histories can.
Examples of coarse-graining are: (1) Specifying the $q^i$ not at
all times but at a discrete set of times. (2) Specifying not all
the $q^i$ at any one time but only some of them. (3) Specifying
not definite values of these $q^i$ but only ranges of values. An
exhaustive set of ranges at one time consists of regions $\{\Delta_\alpha\}$
which make up the whole space spanned by the $q^i$ as $\alpha$ passes
over all values. An exhaustive set of coarse-grained histories
is then defined by sets of such exhaustive ranges $\{\Delta^i_\alpha\}$ at times
$t_i$, $i = 1,\dots, n$.

\begin{figure}[htbp]
  \includegraphics[width=7.5cm]{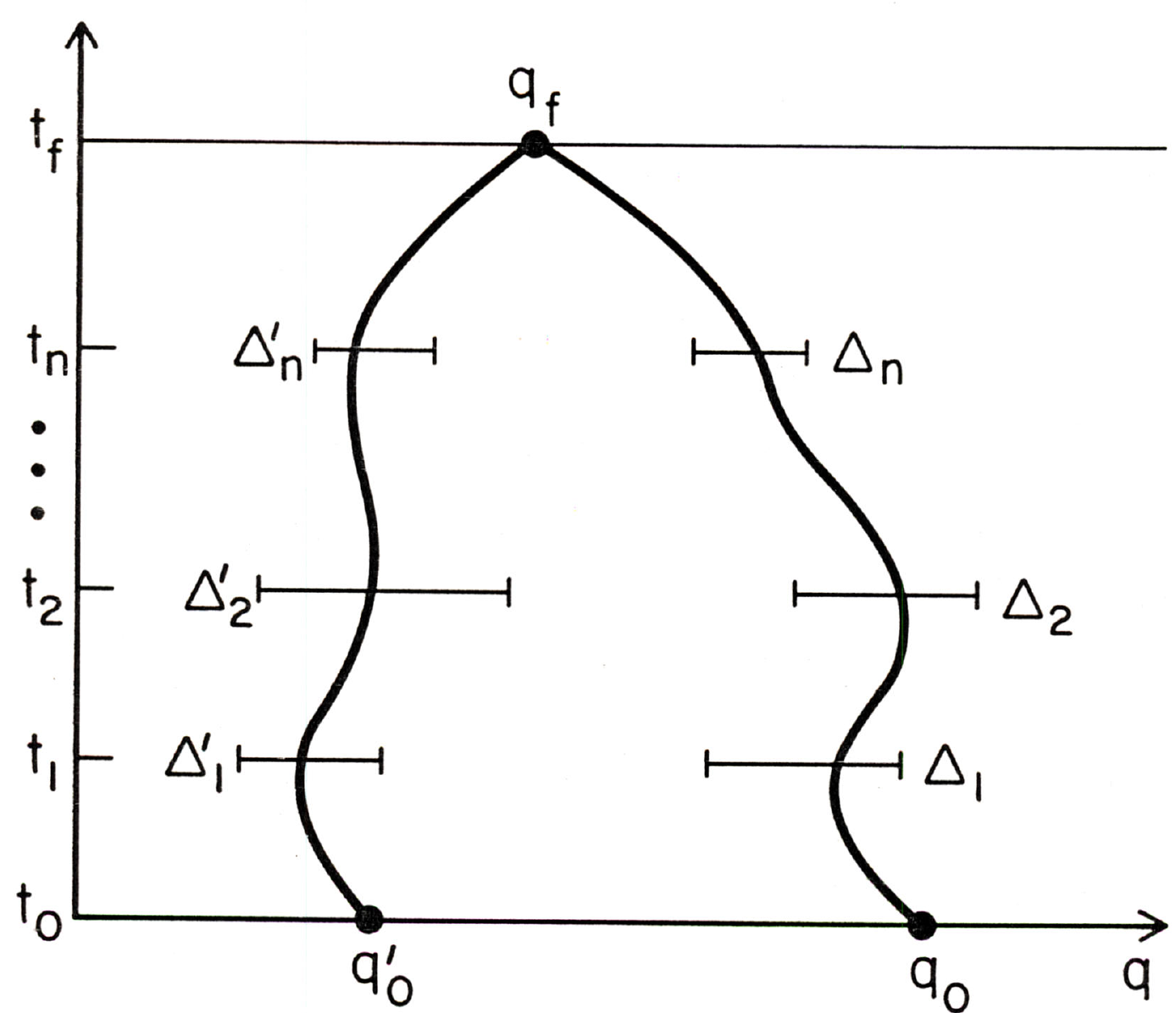}
  \caption{The sum-over histories construction of the decoherence functional.
	\label{fig:2}}
\end{figure}

The important theoretical construct for giving the rule that
determines whether probabilities may be assigned to a given
set of alternative coarse-grained histories, and what these
probabilities are, is the decoherence functional, $D [({\rm history})',
({\rm history})]$. This is a complex functional on
 pairs of histories in the set. In the sum-overhistories
framework for completely fine-grained history
segments between an initial time $t_0$ and a final time $t_f$, it is
defined as follows
\begin{eqnarray}
  D[q^{'i}(t),q^i(t)]&=&\delta(q^{'i}_f-q^i_f) \nonumber \\
	&\times&\exp\left\{i\left(S\left[q^{'i}(t)\right]-
	S\left[q^i(t)\right]\right)/\hbar\right\}
	\rho\left(q^{'i}_0q^i_0\right).
  \label{eq:2.3}
\end{eqnarray}
Here, $\rho$ is the density matrix of the universe in the $q^i$
representation, $q^{'i}_0$ and $q^i_0$ are the initial values of the complete
set of variables, and $q^{'i}_f$ and $q^i_f$ are the final values necessarily
common to both hlstories. The decoherence functional for
coarse-grained histories is obtained from (\ref{eq:2.3}) according to
the principle of superposition by summing over all that is not
specified by the coarse-graining. Thus,
\begin{eqnarray}
  D\left(\left[\Delta_{\alpha'}\right],\left[\Delta_\alpha\right]\right)&=&
  \int_{[\Delta_{\alpha'}]}\delta q'\int_{[\Delta_\alpha]}\delta q\delta(q^{'i}_f-q^i_f) \nonumber \\
	&\times& e^{i\{(S[q^{'i}]-S[q^i])/\hbar\}}\rho(q^{'i}_0,q^i_0).
	\label{eq:2.4}
\end{eqnarray}
More precisely, the integral is as follows (Fig.~\ref{fig:2}): It is over all
histories $q^{'i}(t)$, $q^{'j}(t)$ that begin at $q^{'i}_0$, $q^i_0$ respectively,
pass through the ranges $[\Delta_{\alpha'}]$ and $[\Delta_\alpha]$ respectively,
and wind up at a common point $q^i_f$ at any time $t_f > t_n$. It is completed by
integrating over $q^{'i}_0$, $q^i_0$, $q^t_f$. The three forms of information
necessary for prediction --- state, action, and specific history
are manifest in this formula as $\rho$, $S$, and the sequence of
ranges $[\Delta_\alpha]$ respectively.

The connection between coarse-grained histories and completely
fine-grained ones is transparent in the sum-overhistories
formulation of quantum mechanics. However, the
sum-over-histories formulation does not allow us to consider
coarse-grained histories of the most general type. For the
most general histories one needs to exploit the transformation
theory of quantum mechanics and for this the Heisenberg
picture is convenient. In the Heisenberg picture an exhaustive
set of alternatives at one time corresponds to a set of
projection operators $\{P^k_\alpha(t)\}$,,$\alpha=1,2,3,\cdots$ satisfying
\begin{equation}
  \sum_\alpha P^k_\alpha(t)=1,\ P^k_\alpha(t)P^k_\beta(t)=\delta_{\alpha\beta}P^k_\beta(t).\label{eq:2.5}
\end{equation}
Here, $k$ labels the set of alternatives, $\alpha$ the particular alternative,
and $t$ the time. The operators representing the same
alternatives at different times are connected by
\begin{equation}
  P^k_\alpha(t)=e^{iHt/\hbar}P_\alpha(0)e^{-iHt/\hbar}\label{eq:2.6}
\end{equation}
where $H$ is the total Hamiltonian. Time sequences of such sets
of alternatives define sets of alternative histories for the
univese. A particular history is one particular sequence of
alternatives $(P^1_{\alpha_1}(t_l),P^2_{\alpha_2}(t2),\dots,P^n_{\alpha_n}(t_n))$
corresponding to a particular sequence of $\alpha$'s. We abbreviate such an
individual history by $[P_\alpha]$. With this notation the decoherence functional
in the Heisenberg picture may be written
\begin{equation}
  D([P_{\alpha'}],\ [P_\alpha])={\rm Tr}[P^n_{\alpha'_n}(t_n)\dots
	P^1_{\alpha'_1}(t_1)\rho P^1_{\alpha_1}(t_1)\dots P^n_{\alpha_n}(t_n)],\label{eq:2.7}
\end{equation}

In the Heisenberg picture a completely fine-grained set of
histories is defined by giving a complete set of projections at
each and every time. Every possible set of alternative histories
may then be obtained by coarse graining the various fine-grained
sets, that is by using $P$'s in the coarser grained sets
which are sums of those in the fine-grained sets. Thus, if
$\{[\bar P_\beta]\}$ is a coarse-graining of the set of histories $\{[P_\alpha]\}$,
we write
\begin{equation}
  D([\overline{P_{\beta'}}], [\overline{P_\beta}])=
	\sum_{{\rm all\ }P_{\alpha'}\ {\rm not\ fixed\ by\ }[\overline{P_{\beta'}}]}
	\sum_{{\rm all\ }P_\alpha\ {\rm not\ fixed\ by\ }[\overline{P_\beta}]}
	D([P_{\alpha'}],\ [P_\alpha]).
	\label{eq:2.8}
\end{equation}
A set of coarse-grained alternative histories is said to
decohere when the off-diagonal elements of $D$ are sufficiently
small to be considered vanishing for all practical purposes:
\begin{equation}
  D([P_{\alpha'}],\ [P_\alpha])\approx0,\ {\rm for\ any\ }\alpha'_k\ne\alpha_k.\label{eq:2.9}
\end{equation}
This is a generalization of the condition for the absence of
interference in the two-slit experiment (approximate equality
of the two sides of (\ref{eq:2.2})).

The rule for when probabilities can be assigned to histories
of the universe is then this: To the extent that a set of
alternative histories decoheres, consistent probabilities can be assigned
to its individual members. The probabilities are the diagonal
elements of $D$. Thus,
\begin{eqnarray}
  p([P_\alpha])&=&D([P_\alpha],\ [P_\alpha]) \nonumber \\
	&=&{\rm Tr}[P^n_{\alpha_n}(t_n)\dots P^1_{\alpha_1}(t_1)\rho P^1_{\alpha_1}(t_1)\dots
	P^n_{\alpha_n}(t_n)]\label{eq:2.10}
\end{eqnarray}
when the set decoheres.

The probabilities defined by (\ref{eq:2.10}) obey the rules of
probability theory as a consequence of decoherence. The
principal requirement is that the probabilities be additive on
``disjoint sets of the sample space". For histories this means
the sum rules
\begin{equation}
  p([\overline{P_\beta}])\approx
	\sum_{{\rm all\ }P_\alpha\ {\rm not\ fixed\ by\ }[\overline{P_\beta}]}
	p([P_\alpha]).\label{eq:2.11}
\end{equation}
These relate the probabilities for a set of histories to the
probabilities for all coarser grained sets that can be constructed
from it. For example, the sum rule eliminating all
projections at only one time is, in an obvious notation:
\begin{eqnarray}
  \sum_{\alpha_k}&&p(\alpha_nt_n,\dots\alpha_{k+1}t_{k+1},\alpha_kt_k,\alpha_{k-1}t_{k-1}
	\dots\alpha_1t_1) \nonumber \\
	&\approx&p(\alpha_nt_n,\dots\alpha_{k+1}t_{k+1},\alpha_{k-1}t_{k-1}
	\dots\alpha_1t_1).\label{eq:2.12}
\end{eqnarray}
Given this discussion, the fundamental formula of quantum
mechanics may be reasonably taken to be
\begin{equation}
  D([P_{\alpha'}],\ [P_\alpha])\approx\delta_{\alpha'_1\alpha_1}\dots
	\delta_{\alpha'_n\alpha_n}p([P_\alpha])\label{eq:2.13}
\end{equation}
for all $[P_\alpha]$ in a set of coarse-grained alternative histories.
Vanishing of the off-diagonal elements of $D$ gives the rule for
when probabilities may be consistently assigned. The diagonal
elements give their values.

Decoherent histories of the universe are what we may
utilize in the quantum mechanical process of prediction for
they may be assinged probabilities. Decoherence thus
generalizes and replaces the notion of ``measurement", which
served this role in the Copenhagen interpretations. Decoherence
is a more precise, more objective, more observer-independent idea  and gives a definite meaning to Everett's
branches. If their associated histories decohere, we may assign
probabilities to various values of reasonable scale density
fluctutions in the early universe whether or not anything
like a ``measurement" was carried out on them and certainly
whether or not there was an ``observer" to do it.

\section{Origins of decoherence}\label{sec:3}

The decoherence of a set of alternative histories is not a
property of their coarse-graining alone. As the formula (\ref{eq:2.7})
for $D$ shows, it depends on all three forms of information
necessary to make predictions about the universe and in
particular on its quantum state.  Given $\rho$ and $H$,
we would compute which sets of alternative histories
decohere and there would be a great many such sets.

We are not likely to carry out a calculation of all decohering
sets of alternative histories for the universe anytime in the
near future. It is therefore important to investigate specific
mechanisms for decoherence in more restrictive circumstances.
Specific examples of decoherence have been discussed
by many authors, among them Joos and Zeh \cite{ref:4},
Zurek \cite{ref:5}, Caldeira and Leggett \cite{ref:10}, and Unruh and Zurek
\cite{ref:11}. Typically these discussions have considered coarse
grainings defined by projection operators which project onto
a few particular degrees of freedom of a system while ignoring
the rest. The simplest model consists of a single oscillator
interacting linearly with a large number of others. A coarse
graining is used which follows the coordinate of the distinguished
oscillator and ignores the coordinates of the
others. Let $x$ be the coordinate of the special oscillator, $M$ its
mass, $\omega_R$ its frequency renormalized by its interactions with
the others, and $S_{\rm free}$ its free action. Consider the special case
where the density matrix of the whole system, referred to an
initial time, factors into the product of a density matrix
$\bar\rho(x,y)$ of the distinguished oscillator and another for the rest.
Then, generalizing slightly a treatment of Feynman and
Vernon \cite{ref:12}, we can write the decoherence functional defined
by (\ref{eq:2.4}) for this coarse-graining as
\begin{eqnarray}
  D([\Delta_{\alpha'}],\ [\Delta_\alpha])&=&\int_{[\Delta_{\alpha'}]}
	\delta x'\int_{[\Delta_\alpha]}\delta x\delta(x'_f-x_f) \nonumber \\
	&\times&\exp\{i(S_{\rm free}[x'(t)]-S_{\rm free}[x(t)] \nonumber \\
	&+&W[x'(t),\ x(t)])/\hbar\}\bar\rho(x'_0,\ x_0)\label{eq:3.1}
\end{eqnarray}
The sum over the paths of the rest of the oscillators has been
carried out and is summarized by the Feynman-Vernon
influence functional $\exp(i W[x'(t),\ x(t)])$.

The case when the rest of the oscillators are in an initial
thermal state has been extensively investigated by Caldeira
and Leggett \cite{ref:10}. In the simple limit of a uniform cut-off
continuum of oscillators and in the Fokker-Planck limit of
high temperature, they find
\begin{eqnarray}
  W[x'(t),\ x(t)]&=&-M\gamma\int dt[x'\dot{x}'-x\dot{x}+x'\dot{x}-x\dot{x}'] \nonumber \\
	&+&i\frac{2M\gamma kT}{\hbar}\int dt[x'(t)-x(t)]^2\label{eq:3.2}
\end{eqnarray}
where $\gamma$ summarizes the interaction strengths of the distinguished
oscillator with the rest. The real part of $W$ contributes
dissipation to the equations of motion. The imaginary
part squeezes the trajectories $x(t)$ and $x'(t)$ together,
thereby accomplishing decoherence on the characteristic time
scale
\begin{equation}
  t_{\rm decoherence}\gtwid\frac{1}{\gamma}\left[\left(\frac{\hbar}{\sqrt{2MkT}}\right)
	\cdot\left(\frac{1}{d}\right)\right]^2.\label{eq:3.3}
\end{equation}
As emphasized by Zurek \cite{ref:13}, this squeezing can be very rapid
when compared with characteristic dynamical timescales
($t_{\rm decoherence}/t_{\rm dynamical}\sim10^{-40}$ for typical ``macroscopic" values).
In such models coherent phase information is lost by the
creation of correlations with variables which are then ignored
in the coarse-graining and so summed over in constructing
the decoherence functional.

What such models convincingly show is that decoherence
is frequent and widespread in the universe. Joos and Zeh \cite{ref:4}
calculate that a superposition of two states of a grain of
dust localized at two positions 1~mm apart, is decohered simply by the scattering of the
cosmic background radiation on the timescale of a nanosecond.
So widespread is this phenomena with the quantum 
state  and dynamics of our universe that we may meaningfully
speak of habitually decohering variables such as the
center of mass positions of massive bodies.

\section{quasiclassical realms}\label{sec:4}

As observers of the universe, we deal with coarse-grainings
that are appropriate to our limited sensory perceptions,
extended by instruments, communication, and records, but in
the end characterized by a great amount of ignorance. Yet, we
have the impression that the universe exhibits a finer grained
set of decohering histories, independent of us, defining a 
 ``quasiclassical realm", governed largely by classical
laws, to which our senses are adapted while dealing  with
only a small part of it. No such coarse-graining is determined
by pure quantum theory alone. Rather, like decoherence, the
existence of a quasiclassical realm in the universe must be
a consequence of its quantum state and the dynamical theory
describing its evolution.

Roughly speaking, a quasiclassical realm should be a set
of alternative decohering histories that is maximally refined consistent
with decoherence, with individual histories exhibiting
as much as possible patterns of classical correlation in time.
To make the question of the existence of one or more quasiclassical
realms into a {\it calculable}\/ question in quantum cosmology
we need criteria to measure how close a set of
histories comes to constituting a ``quasiclassical realm".
A quasiclassical realm cannot be a {\it completely}\/ fine-grained
description for then it would not decohere. It cannot
consist {\it entirely}\/ of a few ``classical variables" repeated
over and over because sometimes we may measure something
highly quantum mechanical. These variables cannot be
{\it always}\/ correlated in time by classical laws because sometimes
quantum mechanical phenomena cause deviations from
classical physics. We need measures for maximality and
classicality \cite{ref:8}.

It is possible to give crude arguments for the type of
alternatives we expect to occur over and over again
in a set of histories defining a quasiclassical realm. Such  quantities  are called ``quasiclassical variables". In the earliest instants of the universe the
operators defining spacetime on scales well above the Planck
scale emerge from the quantum fog as quasiclassical \cite{ref:14}. Any
theory of the quantum state  that does not imply this is
simply inconsistent with observation in a manifest way. Then,
where there are suitable conditions of low temperature,
density, etc., various sorts of hydrodynamic variables may
emerge as quasiclassical operators. These are integrals over
suitably small volumes of densities of conserved or nearly
conserved quantities. Examples are densities of energy,
momentum, baryon number, and, in later epochs, of  nuclei and
even chemical species. The sizes of the volumes are limited
above by maximality and are limited below by classicality
because they require sufficient ``inertia" to enable them to
resist deviations from predictability caused by their interactions
with one another, by quantum spreading, and by
the quantum and statistical fluctuations resulting from the
interactions with the rest of the universe that accomplish
decoherence. Suitable integrals of densities of approximately
conserved quantities are thus candidates for habitually
decohering quasiclassical operators. These ``hydrodynamic
variables" {\it are}\/ among the principal variables of classical
physics.

It would be in such ways that the classical realm of
familiar experience could be an emergent property of the
early universe, not generally in quantum mechanics, but as a
consequence of our specific quantum state and the theory of dynamcics
describing evolution.

\section{Generalized quantum mechanics}\label{sec:5}

I would now like to turn to the generalization of quantum
mechanics which  may be needed to resolve the conflict
between the need for a preferred time variable in Hamiltonian
quantum mechanics and the inability of any generally
covariant quantum theory of spacetime to supply one. To
start, it is useful to consider what we might mean most
generally by a quantum mechanical theory \cite{ref:9}. Roughly
speaking, by a quantum mechanics we mean a theory that
admits a notion of fine and coarse-grained histories, the
amplitudes for which are connected by the principle of superposition
and for which there is a rule (decoherence) for when
coarse-grained histories can be assigned probabilities obeying
the sum rules of probability calculus. More precisely, from
the discussion in the preceeding section its possible to
abstract the following three elements of quantum mechanics
in general:
\begin{enumerate}[label={(\arabic*)}]
  \item {\it The fine-grained histories:}\/ The exhaustive sets of fine-grained, alternative histories of the universe that are the
most refined description to which one can contemplate
assigning probabilities.
  \item {\it A notion of coarse graining:}\/ A coarse-graining of an
exhaustive set of histories is a partition of that set into
exhaustive and exclusive coarse-grainedclasses $\{h\}$. Various possible coarse-grained
sets of alternative histories may be constructed by
coarse graining the fine-grained sets or by further coarse
graining of an already coarse-grained set.
  \item {\it A decoherence functional:}\/ The decoherence functional,
$D(h, h')$, is defined on each pair of coarse-grained
histories in an exhaustive set, $\{h\}$, for all possible sets including
the completely fine-grained ones. It must satisfy the
following four properties:
  \begin{subequations}
	\begin{enumerate}[label=(\roman*)]
    \item {\it Hermiticity:}
      \begin{equation}
			  D(h,\ h')=D^*(h',\ h),\label{eq:5.1}
      \end{equation}
    \item {\it Positivity:}
		  \begin{equation}
		    D(h,\ h)\ge0,
		  \end{equation}
		\item {\it Normalization:}
		  \begin{equation}
			  \sum_{h,\ h'}D(h,\ h')=1,
			\end{equation}
		\item {\it The principle of superposition:}\/ If $\{\bar h\}$ is a coarser
		graining of a coarse-grained set $\{h\}$ then the decoherence functional
		for the coarser grained set is related to that of the finer grained set by:

		 \begin{equation}
			  D({\bar h},{\bar h'}) = \sum_{h \in {\bar h}} \sum_{h'\in {\bar h}'}D(h,h') .
			\end{equation}

		\end{enumerate}
	\end{subequations}
\end{enumerate}
These three elements are sufficient for the process of prediction.
Decoherence can be defined and probabilities
assigned according to the fundamental formula
\begin{equation} 
{D(h,h')\approx \delta_{h,h'}p(h).}
\end{equation}

As a consequence of the four requirements (\ref{eq:5.1}) for  the
decoherence functional these probabilities will obey the rules
of probability calculus. With these probabilities the thoery
becomes predictive.

Thus, the three elements of Hamiltonian quantum mechanics 
are as follows: (1) The set of fine-grained histories are defined
by sequences of sets projections onto {\it complete} sets of
states, one set at each time. (2)  Sets of alternative coarse-grained 
histories defined by sequences of  sets of projections which are sums
of projections in finer grained ones. (3) Consistency of the probabilities for histories with the rules of probability theory.
with the rules of probability theory. 

Hamiltonian quantum mechanics
is not the only way of constructing a theory with the three
elements of generalized quantum mechanics. More general
possibilities may be considered, and, as we shall argue below,
may be useful in constructing a generally covariant quantum
theory of spacetime.

\begin{figure}[htbp]
  \includegraphics[height=4in]{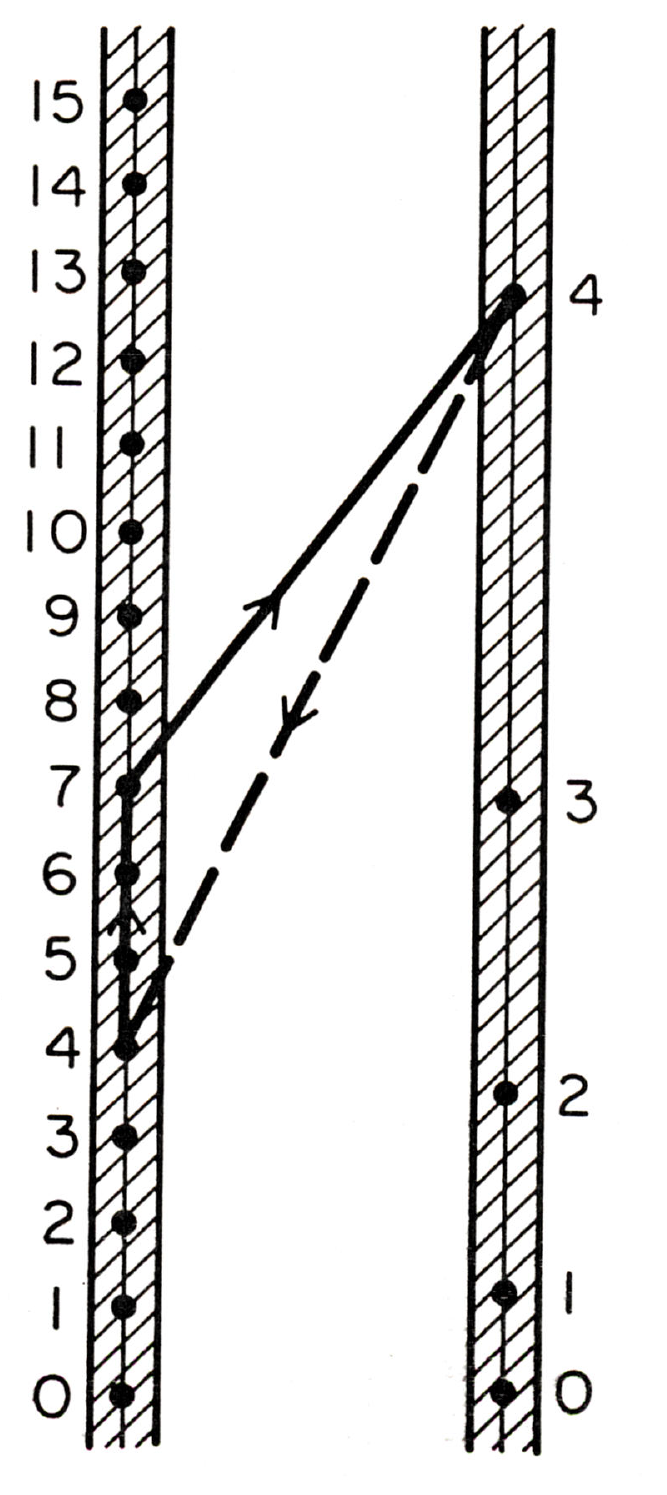}
  \caption{Closed timelike lines in a wormhole geometry. The figure shows a
spacetime diagram (time upward) with two wormhole mouths (the shaded
regions). The wormhole geometry is multiply connected so that it is possible
to pass nearly simultaneously from points in one wormhole mouth to
 the other. The wormhole mouth on the left remains at rest in an inertial frame.
The one at right is initially at rest with respect to the first at $t = 0$ but then
begins to rotate about it. The figure shows the corotating frame and the
readings of a clock at the center of each wormhole mouth. As a consequence
of time dilation in the rotating mouth, this spacetime has closed timelike
curves of which one is shown. The dotted segment represents the nearly
instantaneous passage through the wormhole throat.
	\label{fig:3}}
\end{figure}

An interesting example of a generalized quantum mechanics
is provided by field theory in the kind of wormhole
spacetime discussed by Morris, Thorne,Yurtsever \cite{ref:15} and
others that is illustrated in Fig.~\ref{fig:3}. These are not the
four-dimensional wormholes discussed in connection with the
value of the cosmological constant. They are handles on
three-dimensional space. The topology of spacetime is
${\bf R}\times M^3$ with $M^3$ being multiply connected.

Imagine that before some time $t = t_s$, the wormhole
mouths are at rest with respect to one another. At time $t_s$
they begin to rotate about one another and continue until a
moment of time symmetry when they reverse their motion
eventually coming to relative rest at time $t_e$. Before $t_s$ and
after $t_e$ there are no closed timelike lines and it is possible
to define surfaces of constant time that foliate those portions
of spacetime. In between $t_s$ and $t_e$, however, because of
time dilation in the rotating wormhole mouth there are
closed timelike lines, as Fig.~\ref{fig:3} illustrates. By going through
the wormhole throat it is effectively possible to go backward
in time. Such wormhole spacetimes are time orientable but
not causal.

It is clear that there is no straightforward Hamiltonian
quantum mechanics in a wormhole spacetime between the
surfaces $t_s$ and $t_e$. What would be the surfaces of the preferred
time? How would unitary evolution of arbitrary states in the
Hilbert space be defined in the presence of closed timelike
lines?

A generalized quantum mechanics of the kind we have
been discussing may be constructed for this
example using a sum-over-histories decoherence functional.
The three ingredients in this construction would be the
following:
\begin{enumerate}
  \item {\it Fine-grained histories:}\/ For the fine-grained histories we
may take single-value field configurations, $\phi(x)$, in the wormhole
spacetime.
  \item {\it Coarse-grainings:}\/ The fine-grained histories may be
partitioned according to their values on spacetime regions.
Select a set of  regions, specify an exhaustive set of
ranges for the average values of the field in these regions, and
one has partitioned the four-dimensional field configurations
into classes, $\{h\}$, that have the various possible values of the
average field. For example one might specify the spatial field
configurations on an initial  constant time surface with $t < t_s$
and on a final constant time surface with $t > t_e$. The resulting
probabilties would be relevant for defining the $S$-matrix for
scattering from the wormhole.
 \item {\it Decoherence functional:} A sum-over-histories decoherence
functional is: \end{enumerate}
\begin{eqnarray}
  D(h,\ h')&=&\int_h\delta\phi\int_{h'}\delta\phi'\delta[\phi_f(x)-\phi'_f(x)] \nonumber \\
	&\times&\exp\{i(S[\phi(x)]-S[\phi(x')])/\hbar\}\rho_0[\phi_0(x),\ \phi'_0(x)].\label{eq:5.3}
\end{eqnarray}
Here, the integrations are over single-valued field configurations
between some initial constant time surface $t_0 < t_s$
and some final constant time surface $t_f > t_e$. $\phi_0(x)$ and $\phi'_0(x)$
are the spatial configurations on the initial surface; their integral
is weighted by the density matrix $\rho_0$. $\phi_f(x)$ and $\phi'_f(x)$ are the
spatial configurations on the final surface; their coincidence is
enforced by the functional $\delta$-function. The integral over $\phi(x)$
is over the class of field configurations in the class $h$. For
example, if $h$ specifies the average value of the field is some
region to lie in a certain range, then the integral is only over
$\phi(x)$ that have such average values. Formally, this decoherence
functional satisfies four conditions of eq.~(\ref{eq:5.1}).

With the generalized quantum mechanics based on the three
elements described above probabilities can be assigned to
coase-grained sets of field histories in the wormhole spacetime.
These probabilities obey the standard probability sum rules.
There is no equivalent Hamiltonian formulation of this quantum
mechanics because this wormhole spacetime, with is
closed timelike lines provides no foliating family of spacelike
surfaces to define the required preferred time. Nevertheless, the
generalized theory is predictive. What has been lost in this
generalization is any notion of ``state at a moment of time" and
of its unitary evolution in between the surfaces $t_e$ and $t_s$. This
is perhaps not surprising for a region of spacetime that has no
well defined notion of ``at a moment of time".

\section{A quantum mechanics for spacetime}\label{sec:6}

I  will now  sketch how a generalized quantum
mechanics for spacetime might be constructed that does not
break general covariance by singling out a preferred family of
spacelike surfaces for the distinguished time variable that
would be needed in Hamiltonian formulation of quantum mechanics.  And then I shall  discuss 
 how familiar Hamiltonian quantum mechanics  of quantum mechaniscould be
an emergent approximation to this more general framework appropriate
because of the classical spacetime of the late universe \cite{ref:16}.

\begin{figure}[htbp]
  \includegraphics[width=5.5in]{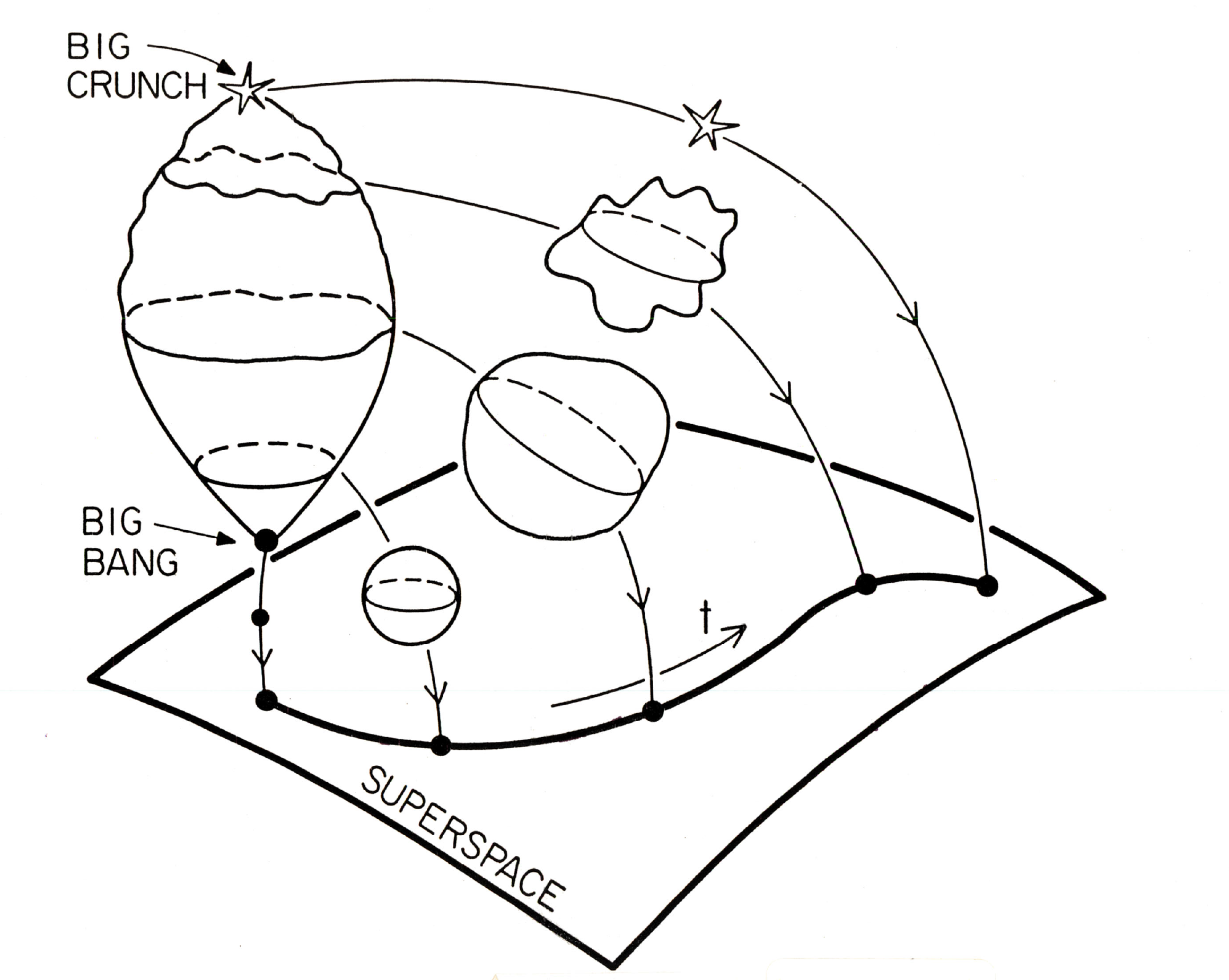}
  \caption{Superspace. A cosmological history is a four-dimensional cosmological
spacetime with matter fields upon it. A two dimensional representation
of such a history is shown in the upper left of this figure proceeding from a big
bang to a big crunch. A cosmological history can be thought of as a succession
of three-dimensional geometries and spatial matter field configuration. Superspace
is the space of such three-dimensional geometries and matter field
configurations. A ``point" in superspace is a particular three-geometry and
spatial matter field configuration. The succession of three-geometries and
matter fields that make up a four-geometry and field history and thereforee therefor trace
out a path in superspace.
	\label{fig:4}}
\end{figure}

In cosmology a history is a  cosmological four-geometry
with a four-dimensional matter field configuration upon it. An
example is the classical Friedman evolution of a closed universe
from the big bang to a big crunch. In quantum mechanics
all the possible histories, classical and non-classical, must be
assigned amplitudes. Classical or non-classical, cosmological
histories may be thought of as successions of three-dimensional
geometries. The Friedman universe is a three-sphere expanding
and contracting according to the Einstein equation. Nonclassical
histories can have arbitrarily varied histories of expansion
and contraction. Thus, cosmological histories may be
thought of as {\it paths}\/ in the superspace of three-geometries and
three dimensional matter configurations (Fig.~\ref{fig:4}).

One natural way of defining a coarse-graining of cosmological
histories is to utilize a family of regions in superspace,
$\{R_\alpha\}$, and partition the paths according to how they
pass through them. A history which passes through, say each
of three regions at least once has at least three spacelike surfaces
with geometries and matter field configurations specified to
an accuracy determined by the sizes of the regions. A {\it set}\/ of
superspace regions defines a {\it partition of the possible spacerimes}\/ into exhaustive and
exclusive  coarse-grained classes. For example, with two regions there is the
class of paths which go through both regions at least once, the
class of paths which go through the first region at least once
but never the second, the class which goes through none of
the regions, and so forth. A sum-over-histories decoherence
functional on the various classes $\{h\}$ defining such a coarse-grained
set is then naturally constructed as follows:
\begin{equation}
  D(h,\ h')=\int_{h,{\cal C}}\delta g\delta\phi\int_{h',{\cal C}}\delta g'\delta\phi'
	\exp\{i(S[g,\phi]-S[g',\phi'])/\hbar\}\label{eq:6.1}
\end{equation}
The sum over $(g, \phi)$ is over histories which start from
prescribed quantum states (say, the ``no boundary"
proposal) and proceed through regions $\{R_\alpha\}$ as specified by
the class $h$ to a final condition representing complete ignorance.
The sum over $(g', \phi')$ is similarly over histories in the
partition $h'$. The density matrix and the final $\delta$-function
which occur in other sum-over-histories expressions like (\ref{eq:2.4})
are here expressed in terms of conditions on the paths, ${\cal C}$.
With appropriate conditions, this construction satisfies the
requirements for a decoherence functional discussed above.
With this decoherence functional the fundamental formula
(5.2) can be used to identify decoherent sets of histories and
assign them probabilities that obey the rules of probability
theory to the level that decoherence is enforced.

There is, in general, no possible choice of time variable
such that this quantum mechanics of spacetime can be put
into the Hamiltonian form. For that to be the case we would
need a time function on superspace whose constant time
surface the histories cross once and only once. There is none.
Put differently, there is no geometrical quantity which
uniquely labels a spacelike hypersurface. The volume of the
universe, for example, may single out just a few surfaces in a
{\it classical}\/ cosmological history, but in quantum mechanics we
must consider all possible histories, and a non-classical history
may have arbitrarily many surfaces of a given volume.

However, while we do not recover a Hamiltonian formulation
precisely and generally we may recover it {\it approximately}\/
for special coarse-grainings in restricted regions of
superspace and for a particular quantum state. Suppose for
example, the the quantum state  was such that for coarse
grainings defined using  sufficiently unrestrictive regions, in a
regime of three-geometries much larger than the Planck scale,
only a single spacetime geometry $\hat g$ contributed to the geometrical
sums  :(6.1) defining the decoherence functional. Then
we would have
\begin{equation}
  D(h,\ h')=\int_{h,{\cal C}}\delta\phi\int_{h',{\cal C}}\delta\phi'
	\exp\{i(S[\hat g,\phi]-S[\hat g,\phi'])/\hbar\}.\label{eq:6.2}
\end{equation}
This decoherence functional defines the quantum mechanics
of a field theory of $\phi$ in the background spacetime $\hat g$. Any
family of spacelike surfaces in this background picks out a
unique field configuration since the sums are over fields which
are single-valued on spacetime. There is a notion of causality,
and we recover a sum-over-histories expression of the field
theory of $\phi$ in the background spacetime $\hat g$ in a particular
quantum state. This does have an equivalent
Hamiltonian formulation.

It could be in this way that the familiar Hamiltonian
framework quantum mechanics emerges an approximation
appropriate to the existence of an approximate classical
spacetime --- an approximation which is not generally valid
in quantum theories, but appropriate to our special place late
in a universe with particular quantum states.

\section*{Conclusions}

In the history of physics, ideas that were once seen to be
fundamental, general, and inescapable parts of the theoretical
framework are sometimes later seen to be consequent,
special, and but one possibility among many in a yet more
general theoretical framework. This is often for the following
reason: The idea was not a truly general feature of the
world, but only {\it perceived}\/ to be general because of our special
place in the universe and the limited range of our observations
\cite{ref:17}. Examples are the earth-centered picture of the
solar system, the Newtonian notion of time, the exact status
of the laws of the thermodynamics, the Euclidean laws of
spatial geometry, and classical determinism. In veiw of this
history, it is appropriate to ask of any current theory ``which
ideas are truly fundamental and which are `excess baggage'
that can be viewed more successfully as but one possibility
out of many in a yet more general theoretical framework?" In
cosmology it is especially appropriate to ask this question.
We live in a special position in the universe, not so much in
place, as in time. As observers of the Universe we are late, living billions years
after the big bang, a time when many interesting possibilties
for physics could be realized which are not easily accessible
now. Moreover, we live in a special universe whose smooth,
perhaps comprehensible, quantum state  is but one of the
many we could imagine.

This article has advanced the point of view that there are
two features of common quantum mechanics usually taken to
be fundamental  may be special, approximate, emergent
features of the late epoch of a universe with our kind of quantum state
These are the ``quasiclassical realm of familiar
experience and the Hamiltonian framework of quantum
mechanics with its preferred time variable. Before the Planck
time there are unlikely to have been classically behaving
variables of any sort. In particular it is unlikely for there to
have been a classically behaving background spacetime to
supply the preferred time of Hamiltonian quantum mechanics.
However, although a ``quasiclassical realm", so
central to the Copenhagen interpretations of quantum
mechanics may not be a general feature of all epochs of all
universes, it may be seen as an approximtate feature of the late
epoch of this universe in a more general post-Everett formulation of quantum mechanics.
Hamiltonian quantum mechanics, with its preferred
time, may not be the most general formulation of quantum
mechanics, but it may be an approximation to a more general
sum-over-histories formulation appropriate to the late
epochs of a universe, like ours, whose intial condition implies
classical spacetime there.

\section*{Acknowledgments}
Preparation of this report as well as the work it describes were supported in
part by NSF grants PHY85-06686, PHY90-0850 and PHY18-18018105. The author thanks Debbie Ceder
for retyping the manuscript.


\end{document}